\documentclass[fleqn]{article}

\usepackage{bcprules}

\usepackage{latexsym}
\usepackage{graphics}
\newcommand{\Calc}[1]{\begin{description}
                \item \begin{tabbing}\qquad\=\quad\=\kill
                \> \> #1\end{tabbing}
                \end{description}}
\newcommand{\conn}[2]{\\*$#1 $\> \{#2\}\\\> \>}

\makeatletter
\newcommand{\calcEqLabel}[1]{%
                                \refstepcounter{equation}%
                                \label{#1}%
                                \rlap{\kern-\@totalleftmargin\kern-2em%
                                     \hbox to\hsize{\hfil(\arabic{equation})}}%
                                \ignorespaces
                             }
\makeatother

\newenvironment{mproof}{\begin{tabbing}
 \q \=xxxxxxxxxxxxxxxxxxxxxxxxx\=xxxx\=xxxx\=xxxx\=xxxx\=xxxx\=xxxx\=xxxx\=xxxx\=xxxx\kill
\+ \kill}{\end{tabbing}}

\newcommand{\mquad}{\qquad}
\newenvironment{mproof2}{\begin{tabbing}
 \mquad \=xxx\=xxx\=xxx\=xxxx\=xxxx\=xxxx\=xxxx\=xxxx\=xxxx\=xxxx\=xxxx\=xxxx\=xxxxxxxxxxxxxxxx\kill
\+ \kill}{\end{tabbing}}

\newcommand {\openproof} [1] {\vspace{2ex} \noindent $\bullet$ {#1} }

\newcommand{\q}{\hspace*{\mathindent}}    
\newcommand{\tb}{\>}       
\newcommand{\ttb}{\> \>}      
\newcommand{\tttb}{\> \> \>}     
\newcommand{\ttttb}{\> \> \> \>}    
\newcommand{\tfiveb}{\> \> \> \> \>}   
\newcommand{\tsixb}{\> \> \> \> \> \>}    
\newcommand{\tsevenb}{\> \> \> \> \> \> \>}  
\newcommand{\tnineb}{\> \> \> \> \> \> \> \> \>}   

\newcommand{\andl}{\land}
\newcommand{\orl}{\lor}
\newcommand{\impl}{\Rightarrow}
\newcommand{\eql}{\equiv}
\newcommand{\notl}{\lnot}

\newcommand{\true}{\mbox{\em true}}
\newcommand{\false}{\mbox{\em false}}

\newcommand{\inference}[2]
{\frac{\raisebox{2pt}{\normalsize $#1$}}
      {\raisebox{-2pt}{\normalsize $#2$}}
}

\newcommand{\lkwd}[1]{{\bf \small #1}\ }

\newcommand{\lrkwd}[1]{{\bf \small #1}}

\newcommand{\en}{\mbox{\ {\bf en}\ }}
\newcommand{\len}{\mbox{{\bf en}\ }}

\newcommand{\lren}{\mbox{{\bf en}}}

\newcommand{\transient}{\mbox{{\bf transient}\ }}
\newcommand{\rtransient}{\mbox{\bf transient}}

\newcommand{\lt}{\mbox{\em leads-to}}
\newcommand{\co}{\mbox{\ {\bf co}\ }}
\newcommand{\lco}{\mbox{{\bf co}\ }}

\newcommand{\lrco}{\mbox{{\bf co}}}

\newcommand{\initially}{\mbox{{\bf initially}\ }}

\newcommand{\constant}{\mbox{{\bf constant}\ }}
\newcommand{\rconstant}{\mbox{\bf constant}}

\newcommand{\stable}{\mbox{{\bf stable}\ }}
\newcommand{\rstable}{\mbox{\bf stable}}

\newcommand{\invariant}{\mbox{{\bf invariant}\ }}
\newcommand{\rinvariant}{\mbox{\bf invariant}}

\newcommand{\ltt}{\ {\boldmath \mapsto}\ }  

\newcommand {\pata}     {mxxx\=xx\=xx\=xx\=xx\=xx\=xx\=xx\=xx\=xx\= \kill}
\newcommand {\pg} [1] {
  \vspace*{-0.2cm}
  \begin{tabbing} \pata \+ \\
    {\bf program} {\em {#1}}$\, ;$ \+ \\
}
\newcommand {\pgend} [1] {
  \-\\ 
  {\bf end} \{{\em {#1}}\}\\ 
  \end{tabbing}
  \vspace*{-0.2cm}
}

\newcommand{\End}{\hfill $\Box$} 

{\subsubsection*{Example} \vspace{-0.05in}}
{\End}

\newenvironment{Example'} 
{\subsubsection*{Example}\vspace{-0.05in}}
{}

\newcommand{\ra}{\ \ \mbox{\boldmath{$\to$}}\ \ }

\newenvironment{gtab}{\begin{tabbing}
    \mquad\=xx\=xx\=xx\=xx\=xx\=xx\=xx\=xx\=xx\=xx\=xx\=xx\=xx\ \kill
    \+ \kill}{\end{tabbing}}

\newenvironment{gtab5}{\begin{tabbing}
    \=xx\=xx\=xx\=xx\=xx\=xx\=xx\=xx\=xx\=xx\=xx\=xx\=xx\ \kill
    \+ \kill}{\end{tabbing}}

\newcommand{\HIDE}[1]{}

\newtheorem{Sublemma'}{Sublemma}

\newtheorem{Lemma'}{Lemma}

\newtheorem{Theorem'}{Theorem}

\newtheorem{Corollary'}{Corollary}

\newtheorem{Observation'}{Observation}

\newtheorem{Proposition'}{Proposition}

\newtheorem{Subproposition'}{Subproposition}

\usepackage{amsmath}
\usepackage{listings}

\title{Notes on Orc}

\newcommand{\otw}{\ \mbox{\boldmath{$;$}}\ } 
\newcommand{\bnfbar}{\;\,{\mbox{\rule[-0.8mm]{0.38mm}{3.6mm}}}\;\,}

\renewcommand{\tt}{\ttfamily\small}
\lstdefinelanguage{orc}{
  morekeywords={val,def,as,include,type,site,class,stop,if,then,else,Ift,Iff,signal,lambda,atomic},
  sensitive=false,
  morecomment=[l]--,
  morecomment=[n]{\{-}{-\}},
  morestring=[b]",
}
\newcommand{\vb}[1]{\lstinline!#1!}

\usepackage{amssymb}

\begin{document}

\title{Bilateral Proofs of Safety and Progress Properties of
  Concurrent Programs}

\author{Jayadev Misra\\ University of Texas at Austin, misra@utexas.edu}

\date{April 5, 2017}

\maketitle

\begin{abstract}
This paper suggests a theomisra@utexas.edury of composable specification of
concurrent programs that permits: (1) verification of program
code for a given specification, and (2) composition of the
specifications of the components to yield the specification of a
program. The specification consists of both terminal properties
that hold at the end of a program execution (if the execution
terminates) and perpetual properties that hold throughout an
execution. We devise (1) proof techniques for verification,
and (2) composition rules to derive the specification of a
program from those of its components. We employ terminal
properties of components to derive perpetual properties of a
program and conversely. Hence, this proof strategy is called
bilateral. The compositional aspect of the theory is important in
assembling a program out of components some of whose source code
may not be available, as is increasingly the case with
cross-vendor program integration.\\

\noindent
Keywords: Program specification,
  concurrent programming,
  verification,
  composition,
  safety and progress properties.
\end{abstract}


\newcommand{\fbb}{[\! \,]}
\newcommand{\raa}{\!\rightarrow\!}
\newcommand{\uk}{_{k+1}}
\newcommand{\postf}{\mathit{post_f}}

\section{Introduction}
Four decades of intensive research has failed to yield a scalable
solution to the problem of concurrent program design and
verification. While there have been vast improvements in our
understanding, the theory and practice in this area lag considerably
behind what has been achieved for sequential programs. Very small
programs, say for synchronization, are proved manually, though the proof
methods are mostly unscalable. Larger programs of practical
significance, say cache coherence protocols, are typically proved
using model checking, which imposes size limitations. Programs from
different vendors are rarely assembled to run concurrently.

We believe that the problem stems from the lack of a theory of
\emph{composable specification} for concurrent programs. Sequential
imperative programs enjoy such a theory, introduced by
Hoare~\cite{Hoare69}, in which a program is specified by a pair of
predicates, called its pre- and postcondition. The theory successfully
permits: (1) verification of program code for a given specification,
and (2) composition of the specifications of the components to yield
the specification of a program. A fundamental concept is
\emph{invariant} that holds at specific program points, though
invariant is not typically part of the program specification.
Termination of a program is proved separately.

A specification of a concurrent program typically includes not just
the pre- and postconditions but properties that hold of an entire
execution, similar to invariants. A typical specification of a thread
that requests a resource, for example, may state that: (1) once it
requests a resource the thread waits until the resource is granted,
and (2) once the resource is granted the thread will eventually
release it. The first property is an instance of a \emph{safety}
property and the second of a \emph{progress} property, see
Lamport~\cite{Lamport} and Owicki and Lamport~\cite{O+L}.

Call the postcondition of a program for a given precondition to be a
\emph{terminal} property. And a property that holds throughout an
execution a \emph{perpetual} property. Terminal properties compose
only for sequential programs, though not for concurrent programs, and
they constitute the essence of the assertional proof method of
Hoare. Safety and progress are typical perpetual properties.

This paper suggests a theory of composable specification of concurrent
programs with similar goals as for sequential programs. The
specification consists of both terminal and perpetual properties. We
devise (1) proof techniques to verify that program code meets a given
specification, and (2) composition rules to derive the specification
of a program from those of its components. We employ terminal
properties of components to derive perpetual properties of a program
and conversely. Hence, this proof strategy is called
\emph{bilateral}. The compositional aspect of the theory is important
in assembling a program out of components some of whose source code
may not be available, as is increasingly the case with cross-vendor
program integration.

The Hoare-proof theory for sequential programs is known to be sound
and relatively-complete. A sound and relatively-complete proof theory
for concurrent programs that use a very limited set of program
constructs, known as \emph{Unity}, appears in Chandy and
Misra~\cite{CM88} and Misra~\cite{seuss}. This paper combines both
approaches to yield a theory applicable to concurrent programs written
in conventional languages. (The soundness and relatively-completeness
of this theory has not been proved yet.)

We treat three examples of varying complexity in detail in this
paper. First, a program that implements a distributed counter is used
to illustrate the various concepts of the theory in stages through out
the paper. Appendix~\ref{Mutex} includes a proof of a mutual exclusion
program, a system of tightly-coupled components. Unlike traditional
proofs, it is based on composing the specifications of the components.
Appendix~\ref{CAfold} includes proof of a recursively-defined
concurrent program, where code of one component is only implicit,
again using composable specifications. We know of no other proof
technique that can be used to prove this example program.

\section{Program and Execution Model}
\subsection{Program Structure}
The syntax for programs and its components is given below. 

\begin{gtab}
\tttb  \emph{action}  \tfiveb ::= guard $\ra$ body \\
$f,\ g$\ :: \tttb \emph{component} \tfiveb ::= \emph{action}\ $\vert$\ $f\ \fbb\ g$
\ $\vert$\ seq$(f_0,\ f_1, \cdots f_n)$\\
\tttb  \emph{program}  \tfiveb ::= $f$
\end{gtab}

An action has a guard, which is a predicate, and a body which is a
piece of code. Execution of the body in a state in which the guard
holds is guaranteed to terminate; we assume that this guarantee is
independently available. Details of execution semantics is given
Section~\ref{Program execution}. A guard that is $\true$ is often
dropped. An action without a guard is \emph{non-blocking} and a
guarded action \emph{blocking}.

A structured component is either: (1) a \emph{join} of the form
$f\ \fbb\ g$ where $f$ and $g$ are its \emph{direct subcomponents}, or
(2) seq$(f_0,\ f_1, \cdots f_n)$ where the direct subcomponents,
$f_i$, are combined using some sequential language construct. A join
models concurrent execution, as we describe in Section~\ref{Program
  execution}. And seq denotes any sequential language construct for
which proof rules are available. Thus, the typical constructs of
sequencing, if-then-else and do-while are seq constructs. A
subcomponent of any component is either its direct subcomponent or a
subcomponent of some direct subcomponent. Note that a component is
never empty.

Join construct is commutative and associative.  The syntax permits
arbitrary nesting of sequential and concurrency constructs, so,
``$(f\ \fbb\ g) \otw (f'\ \fbb\ g')$'' is a program with $f,\ f',\ g$
and $g'$ as components.

A \emph{program} is a component that is meant to be executed alone.  

\paragraph{Access rights to variables} 
A variable named in a component is \emph{accessible} to it. Variable
$x$ is \emph{local} to $f$ if $f$ has exclusive write-access to $x$
during any of its executions. Therefore, any accessible variable of
a component in a sequential program is local to it. However, a local
variable of $f\ \fbb\ g$ may not be local to either $f$ or $g$.
An accessible non-local variable $x$ of $f$ is \emph{shared}; $x$ is
shared with $g$ if $x$ is also accessible to $g$. Note that $x$ is
local to $f$ and shared in $g$ if $f$ has exclusive write-access to
$x$ and $g$ can only read $x$.

A \emph{program} is executed alone, so, all its accessible variables
are local to it.

A \emph{local predicate} of $f$ is a predicate in which every variable
is a local variable of $f$. Therefore, $\true$ and $\false$ are local
to all components. It follows that the value of a local predicate of
$f$ remains unchanged in any execution as long as $f$ does not take a
step.

\subsection{Program execution}\label{Program execution}
A \emph{step} is an instance of the execution of an action. A step of
action $b\raa \alpha$ is executed as follows: evaluate $b$ without
preemption and if it is true, immediately execute $\alpha$ to
completion, again without preemption ---this is called an \emph{effective
  execution} of the action--- else the program state (and its control point) are
unaltered ---this is called an \emph{ineffective execution}.  Thus, in an
effective execution of $b\raa \alpha$, $b$ holds when the execution of
$\alpha$ begins. Ineffective execution models busy-waiting whereby the
action execution is retried at some future moment.

Execution of a join starts simultaneously for all its direct
subcomponents, and terminates when they all do. At any moment during
the execution, the program control may reside simultaneously in many
of its subcomponents. Execution rules for a seq are the traditional
ones from sequential programming, which ensures that the program
control is within one direct subcomponent at any moment. The
subcomponent itself may be a join which may have multiple control
points of its own.

Initially the program control is at the entry point of the program. In
any given state the program takes a step by choosing a control point
before an action and executing, effectively or ineffectively, the
action. If there is no such control point, the program has
terminated. The choice of control point for a step is arbitrary, but
subject to the following fairness constraint: every control point is
chosen eventually for execution, so that no component is ignored
forever. This is the same constraint advocated by Dijkstra~\cite{D68}
in his classic paper. In contrast to a terminated execution a
deadlocked execution attempts executions of certain actions
ineffectively forever, and the control resides permanently at the
points preceding each of these actions.

An execution is a sequence of steps that can not be extended. Infinite
executions do not have an end state. A finite execution either
terminates or is deadlocked, and it has an end state. It simplifies
the proof theory considerably to imagine that every execution is
infinite, by extending each finite execution by an infinite number of
stutter steps that repeat the end state forever.

\subsection{Example: Distributed counter}\label{counter}
We consider an abstraction of a distributed counter protocol in
Blumofe~\cite{Blumofe}. The proof in this paper is based on a proof by
Rajeev Joshi and the one given in Chapter 6 of Misra~\cite{seuss}.

The protocol $f$ that implements counter $ctr$ is the join of a finite
number of threads, $f_j$, given below. Each $f_j$ has local variables
$old_j$ and $new_j$. Below, each assignment statement and guarded
statement is an action. The following form of the \lrkwd{if} statement
was introduced by Dijkstra~\cite{bib:DijkstraGuarded}; in each
execution of the statement an alternative whose guard holds is
executed without preemption.

\begin{gtab5}
$\initially\ ctr = 0$ \\
$f_j::$\\
\tb$\initially\ old_j,\ new_j = 0,\ 0$ \\ 
\tb\lkwd{loop}\\
\tttb $new_j := old_j+1;$\\
\tttb $\lkwd{if} [\ \ ctr = old_j\ \raa\  ctr := new_j$\\
  \ttttb$\ \vert\ ctr \ne old_j\ \raa\ old_j := ctr\ ]$\\
\tb\lkwd{forever}
\end{gtab5}

It is required to show that $ctr$ behaves as a counter, that is: (1)
$ctr$ never decreases and it increases only by 1, and (2) $ctr$
increases eventually. Both of these are perpetual properties. There is
no terminal property of interest since the program never terminates.

\section{Introduction to the proof theory}
\subsection{Specification}
A specification of component $f$ is of the form
$\{r\}\ f\ \{Q\ \bnfbar\ s\}$ where $r$ and $s$ are the \emph{pre- and
  postconditions}, and $Q$ is a set of \emph{perpetual} properties. The
meaning of such a specification is as follows. For any execution of
$f$ starting in an $r$-state,

\begin{enumerate}
\item if the execution terminates, the end state is an $s$-state, and

\item every property in $Q$ holds for the execution.

\end{enumerate}

We give proof rules for pre- and postconditions in the following
section. Proof rules for perpetual properties appear in subsequent
sections, for safety properties in Section~\ref{Safety Properties} and
for progress properties in Section~\ref{Progress Properties}.

\paragraph{Terminology}
Write $\{r\}\ f\ \{s\}$ when $Q$ is irrelevant in the discussion, and
$\{r\}\ f\ \{Q\ \bnfbar\ \}$ when $s$ is irrelevant. Further for $q\in Q$,
write ``$q \ \mbox{in}\ f$'' when $r$ is understood from the context, and
just ``$q$'' when both $f$ and $r$ are understood. An inference rule
without explicit mention of $f$ or $r$ denotes that the rule applies
for any $f$ and $r$.

\paragraph{Variables named in properties}
A property of component $f$ includes predicates that name accessible
variables of $f$, and other variables, called \emph{free} variables. A
property is implicitly universally quantified over its free
variables. Any inaccessible variable named in a property denotes a
free variable, and, being a bound variable, may be renamed.

\subsection{Local Annotation}
\emph{Local Annotation} of component $f$ associates assertions with
program points such that each assertion holds whenever program control
reaches the associated point in every execution of \emph{any program
  in which $f$ is a component}. Thus, a local annotation yields
precondition for the execution of each action of $f$, and valid pre-
and postcondition of $f$ in any environment. Since the execution
environment of $f$ is arbitrary, only the predicates that are local to
$f$ are unaffected by executions of other components. Therefore, a
local annotation associates predicates local to \emph{each point} of
$f$, as explained below.

Local annotation is defined by the program structure. First, the proof
rule for an action is as follows:\\

\quad $\inference
{\{p\andl b\}\ \alpha\ \{q\}
 }
 {\{p\}\ b\raa \alpha\ \{q\}
 }\\
$

To construct a local annotation of $f =\ $seq$(f_0,\ f_1, \cdots
f_n)$, construct local annotation of each $f_i$ using only the local
variables of $f_i$ as well as those of $f$. Then construct an
annotation of $f$ using the proof rules for seq from the sequential
program proof theory. Observe that the local variables of $f$ are
local to each $f_i$ because in a sequential execution among the direct
subcomponents of $f$ each $f_i$ has exclusive write-access to these
variables during its execution.

To construct a local annotation of $f = g\ \fbb\ h$, construct local
annotations of each of $g$ and $h$ using only their local
variables. Then construct an annotation of $f$ using the proof rule
given below. Note that the proof rule is valid only because the
assertions in $g$ and $h$ are local to those components. \\

\quad $\inference {
  \begin{array}{ccc}
   \{r\}\ g\ \{s\}\\
   \{r'\}\ h\ \{s'\}
  \end{array}
 }
 {\{r\andl r'\}\ g\ \fbb\ h\ \{s\andl s'\} 
 }\\
$

Observe that a local variable of $f$ is not necessarily local to $g$
or $h$ unless they have exclusive write-access to it. Henceforth, all
annotations in this paper are local annotations.

A shortcoming of local annotation is that a variable that is local to
$f\ \fbb\ g$ but shared by both $f$ and $g$ can not appear in a
local annotation by the application of these rules alone. The
invariance meta-rule, given in Section~\ref{meta-rules},
overcomes this problem.

\subsection{Example: Distributed Counter, contd.}\label{dc-contd}
Construct a local annotation of $f_j$ for the example of
Section~\ref{counter}.  Below, we have labeled the actions to refer
to them in subsequent proofs.

\begin{gtab5}
$f_j::$\\
\tb$\initially\ old_j,\ new_j = 0,\ 0$ \\ 
\tb$\{\true\}$\\
\tb\lkwd{loop}\\
\ttb $\{\true\}$\\
\tttb $\alpha_j::\ new_j := old_j+1;$\\
\ttb $\{new_j = old_j+1\}$\\
\tttb $\lkwd{if} [\ \ \beta_j::\ \{new_j = old_j+1\}\ ctr =
  old_j\ \raa\  ctr := new_j\ \ \{\true\}$\\
\ttttb$\ \vert\ \gamma_j::\ \{new_j = old_j+1\}\ ctr \ne
  old_j\ \raa\ old_j := ctr\ \ \ \{\true\}$\\
\ttttb$]$\\
  \ttb $\{\true\}$\\
\tb\lkwd{forever}
\end{gtab5}

\newpage
\subsection{Meta-rules}\label{meta-rules}
The following general rules apply for specifications. 

\begin{itemize}
\item (lhs strengthening; rhs weakening)\\[0.08in]
\quad
$\inference
 {
  \begin{array}{ccc}
    \{r\}\ f\ \{Q\ \bnfbar\ s\}\\
    r' \impl r ,\ s\impl s',\ Q'\subseteq Q,\  \  r' \ \mbox{and $s'$ are
      local to $f$}\ 
  \end{array}
 }
 {\{r'\}\ f\ \{Q'\ \bnfbar\ s'\} 
 } 
 $
 
\item (Conjunction; Disjunction)\\[0.08in]
\quad 
$\inference
 {
  \begin{array}{ccc}
    \{r\}\ f\ \{Q\ \bnfbar\ s\}\\
    \{r'\}\ f\ \{Q'\ \bnfbar\ s'\}
  \end{array}
 }
 {
  \begin{array}{ccc}
    \{r\andl r'\}\ f\ \{Q\cup Q'\ \bnfbar\ s\andl s'\} \\
    \{r\orl r'\}\ f\ \{Q\cap Q'\ \bnfbar\ s\orl s'\}
  \end{array}
 } 
 $

\end{itemize}

\paragraph{Justifications for the meta-rules}
The lhs strengthening and rhs weakening rules are inspired by similar
rules for Hoare-triples. Additionally, since the properties in $Q$ are
independent, any number of them may be removed.

For the conjunction rule, let the set of executions of $f$ starting in
$r$-state be $r$-executions, and, similarly $r'$-executions. The
$r\andl r'$-executions is the intersection of $r$-executions and
$r'$-executions. Therefore, the postcondition of any execution in
$r\andl r'$-executions satisfies $s\andl s'$ and every property in $Q$
or $Q'$, justifying the conjunction rule. The arguments for the
disjunction rule are similar.

\section{Safety Properties}\label{Safety Properties}
A safety property is perpetual. We develop a safety property, $\lrco$,
and its special cases, taken from Misra~\cite{seuss}.  Safety
properties are derived from local annotations and/or safety properties
of the subcomponents of a component. 

\subsection{Safety Property $\lco$}\label{co}
Write $p \co q$ in component $f$, for predicates $p$ and $q$ that may
not be local to $f$, to mean that \emph{effective execution} of any
action of $f$ in a $p$-state establishes a $q$-state. Observe that an
ineffective execution preserves $p$. Thus, given $p \co q$: (1) in any
execution of $f$ once $p$ holds it continues to hold until $q$ is
established, though $q$ may never be established, and (2) as a
composition rule, $p \co q$ holds in component iff it holds in every
subcomponent of it. 

For an annotated component, $\lrco$ is defined by the following proof
rule. \\

$\inference
 {
   \begin{array}{ccc}
     \{r\}\ f\ \{s\} \\
       \ \mbox{For every action}\  b\raa \alpha \ \mbox{with
         precondition}\  pre \ \mbox{in the annotation}:\\
    \{pre\andl b\andl p\}\ \alpha\ \{q\}
  \end{array}
 }
 {\{r\}\ f\ \{p \co q\ \bnfbar\ s\}
 }
$

As an example, the statement ``every change in integer variable
$ctr$ can only increment its value'' may be formalized as $ctr = m \co
ctr = m\ \orl\ ctr = m+1$, for all integer $m$.

\subsection{Special cases of \lco}
Define special cases of $\lrco$ for component $f$: $\rstable$,
$\rconstant$ and $\rinvariant$. Given predicate $p$ and expression $e$,
in any execution of $f$: (1) $\stable p$ means that $p$ remains $\true$ once
it becomes $\true$, (2) $\constant e$ that the value of $e$ never
changes, and (3) $\invariant p$ that $p$ is always $\true$, including
after termination, if the program terminates. Formally, in $f$

\begin{gtab}
  $\stable p$ \tsixb $\eql\ p \ \co\ p$\\
  $\constant e$ \tsixb $\eql\ (\forall c::\ \stable e=c)$\\
  $\invariant p$  \tsixb $\eql\ \initially p$ and $\stable p$
\end{gtab}

Observe that $\invariant \true$ (hence, $\stable \true$) and $\stable
\false$ are properties of every component. A variable for which $f$ has
no write-access is constant in $f$, and so is any expression constructed
out of such variables.

Derived rules for $\lrco$ and some of its special cases, which are
heavily used in actual proofs, are given in
Appendix~\ref{co-derived}. It follows from the derived rules that a
safety property of a program is a property of all its components, and
conversely, as given by the inheritance rule below.

\subsection{Meta-rules}\label{addl-Meta-rules}
\begin{enumerate}
\item (Inheritance) If any safety property ($\lrco$ or any of its
  special cases) holds in all subcomponents of $f$ then it holds in
  $f$. More formally, for safety properties $\sigma$, \\[0.08in]
\begin{tabular}{lr}
\begin{minipage}{.45\textwidth}
\quad Given:\  $\inference
 {(\forall i::\ \{r_i\}\ f_i\  \{s_i\})
 }
 {\{r\}\ f\ \{s\}
 }
$
\end{minipage}

&
\begin{minipage}{.45\textwidth}
\quad Assert:\ $\inference
 {(\forall i::\ \{r_i\}\ f_i\ \{\mathit{\sigma}\ \bnfbar\ s_i\})
 }
 {\{r\}\ f\ \{\mathit{\sigma}\ \bnfbar\ s\}
 }
$

\end{minipage}
\end{tabular}
  
\item (Invariance) A local invariant of a component, i.e., a local
  predicate that is invariant in the component, can be substituted for
  $\true$, and vice versa, in any predicate in an annotation or
  property of the component. In particular, for a program all
  variables are local, so any invariant can be conjoined to an
  assertion including the postcondition.

\end{enumerate}

\paragraph{Justifications for the meta-rules}
Inheritance rule is based on the fact that if a safety property holds
for all components of $f$ it holds for $f$ as well.  Given the 
proof rule at left the inheritance proof rule at right can be asserted
for any set of safety properties $\mathit{\sigma}$.

The invariance rule is from Chandy and Misra~\cite{CM88} where it is
called the ``substitution axiom''. One consequence of the rule is that
a local invariant of $f\ \fbb\ g$, that may not be a local predicate
of either $f$ or $g$, could be conjoined to predicates in an
annotation of $f\ \fbb\ g$. Additionally, all variables in a program
are local; so, any invariant can be substituted for $\true$ in a
program.

\subsection{Example: Distributed Counter, contd.}\label{dc-contd1}
For the example of Section~\ref{counter} we prove that $ctr$ behaves
as a counter in that its value can only be incremented, i.e., $ctr = m
\co ctr = m\orl ctr = m+1$ in $f$. Using the inheritance rule, it is
sufficient to prove this property in every component $f_j$. In $f_j$,
only $\beta_j$ may change the value of $ctr$; so  we need only
show the following whose proof is immediate:

\begin{gtab5}
$\{ctr = m \andl new_j = old_j+1\andl ctr=old_j\}\ ctr := new_j\ \{ctr = m\orl ctr = m+1\}$
\end{gtab5}

\section{Progress Properties}\label{Progress Properties}
We are mostly interested in progress properties of the form ``if predicate
$p$ holds at any point during the execution of a component, predicate
$q$ holds eventually''. Here ``eventually'' includes the current and
all future moments in the execution. This property, called
\emph{leads-to}, is defined in Section~\ref{Leads-to}
(page~\pageref{Leads-to}). First, we introduce two simpler progress properties,
\emph{transient} and \emph{ensures}. Transient is a fundamental
progress property, the counterpart of the safety property
$\rstable$. It is introduced because its proof rules are easy to
justify and it can be used to define ensures. However, it is rarely
used in program specification because ensures is far more useful in
practice. Ensures is used to define $\lt$.

\subsection{Progress Property: transient}\label{transient}
In contrast to a stable predicate that remains true once it becomes
true, a transient predicate is guaranteed to be falsified
eventually. That is, predicate $p$ is transient in component $f$
implies that if $p$ holds at any point during an execution of $f$,
$\notl p$ holds then or eventually in that execution. In temporal
logic notation $p$ is transient is written as $\Box \Diamond \notl
p$. Note that $\notl p$ may not continue to hold after $p$ has been
falsified. Predicate $\false$ is transient because $\false \impl
\true$, and, hence $\notl \false$ holds whenever $\false$ holds. Note
that given $p$ transient in $f$, $\notl p$ holds at the termination
point of $f$ because, otherwise, $f$ can take no further steps to
falsify $p$.

The proof rules are given in Figure~\ref{TransientRules}. Below,
$\postf$ is a local predicate of $f$ that is initially $\false$ and
becomes $\true$ only on the termination of $f$. Such a predicate
always exists, say, by encoding the termination control point into
it. For a non-terminating program, $\postf$ is $\false$. Proof of
$\postf$ is a safety proof.

\newcommand{\dbbb}{\vb{--------------------------------------------------------------------------------}}
\newcommand{\deee}{\vb{--------------------------------------------------------------------------------}}
\begin{figure}[h]\label{TransientRules}\small
\dbbb
\begin{itemize}
\item (Basis)

$\inference
 {
   \begin{array}{ccc}
     \{r\}\ f\ \{s\} \\
       \ \mbox{For every action}\  b\raa \alpha \ \mbox{of $f$ with
         precondition}\  pre:\\
       pre\andl p \impl b\\
       \{pre\andl p\}\ \alpha\ \{\notl p\}\\
  \end{array}
 }
 {
  \{r\}\ f\ \{\transient p\andl \notl \postf\ \bnfbar\ s\}
 }
 $\\

\item (Sequencing)

$\inference
 {
   \begin{array}{ccc}
     \{r\}\ f\ \{\transient p\andl \notl \postf\ \bnfbar\ \postf\}\\
     \{\postf\}\ g\ \{\transient p\ \bnfbar\ \}
   \end{array}
 }
 {
  \{r\}\ f\otw g\ \{\transient p\ \bnfbar\ \}
 }
$\\

\item (Concurrency)

$\inference
 {
  \{r\}\ f\ \{\transient p\ \bnfbar\ \}
 }
 {
  \{r\}\ f\ \fbb\ g\ \{\transient p\ \bnfbar\ \}
 }
$\\

\item (Inheritance)

\begin{tabular}{ll}
\begin{minipage}{.45\textwidth}
    
Given:\ $\inference
 {(\forall i::\ \{r_i\}\ f_i\  \{s_i\})
 }
 {\{r\}\ f\ \{s\}
 }
$
\end{minipage}

&
\begin{minipage}{.45\textwidth}
Assert:\ $\inference
 {(\forall i::\ \{r_i\}\ f_i\ \{\transient p\ \bnfbar\ s_i\})
 }
 {\{r\}\ f\ \{\transient p\ \bnfbar\ s\}
 }
$
\end{minipage}
\end{tabular}

\end{itemize}
\caption{Definition of $\transient$}
\deee
\end{figure}
 

\paragraph{Justifications}
The formal justification is based on induction on the program
structure: show that the basis rule is justified and then inductively
prove the remaining rules. We give informal justifications below.

In the basis rule the hypotheses guarantee that each action of $f$ is
effectively executed whenever $p$ holds, and that the execution
establishes $\notl p$. If no action can be executed effectively,
because precondition of no action holds, the program has terminated
and $\postf$ holds. Hence, $\notl p\orl \postf$, i.e.$\notl (p\andl
\notl \postf)$, hold eventually in all cases.  Note that if $pre \impl
\notl p$ then $pre\andl p$ is $\false$ and the hypotheses are
vacuously satisfied. If $f$ never terminates, $\notl \postf$ always
holds and $\notl p$ is guaranteed eventually.

The next two rules, for sequential and concurrent composition, have
weaker hypotheses. The sequencing rule is based on an observation
about a sequence of actions, $\alpha;\ \beta$. To prove $\transient p$
it is sufficient that $\alpha$ establish $\notl p$ \emph{or that it
  execute effectively, thus establishing $post_{\alpha}$}, and that
$\beta$ establish $\notl p$. The sequencing rule generalizes this
observation to components. Being a local predicate, $\postf$ can not
be falsified by any concurrently executing component, so it holds as
long as the control remains at the termination point of $f$.

In the concurrency rule, as a tautology $g$ either establishes $\notl
p$ eventually, thus guaranteeing the desired result, or preserves
$p$ forever. In the latter case, $f$ establishes $\notl p$ since $\transient
p$ holds in $f$.

The inheritance rule applies to a program with multiple
components. It asserts that if the property holds in each
component $f_i$ then it holds in program $f$. To see this consider two
cases: $f$ is seq or join, and argue by induction on the program
structure.

For seq $f$: if $p$ holds at some point before termination of $f$ it
is within exactly one direct subcomponent $f_i$, or will do so without
changing any variable value. For example, if control
precedes execution of ``\lkwd{if$_b$} \lrkwd{then} $f_0$ \lrkwd{else}
$f_1$'' then it will move to a point preceding $f_0$ or $f_1$ after
evaluation of $b$ without changing the state. Note that $f_i$ may be
a join, so there may be multiple program points within it where
control may reside simultaneously, but all controls reside within one
direct subcomponent of seq $f$ at any moment.  From the hypothesis,
that component, and hence, the program establishes $\notl p$ eventually.

For a join, say $f\ \fbb\ g$: Consider an execution in which, say, $f$
has not terminated when $p$ holds. From the arguments for the
concurrency rule, using that $\transient p$ in $f$, eventually $\notl
p$ is established in that execution. Similar remarks apply for all
executions in which $g$ has not terminated. And, if both $f$ and $g$
have terminated, then $\notl p$ holds from the definition of
$\transient$ for each component. \hfill $\Box$

\paragraph{Notes}
\begin{enumerate}
\item The basis rule by itself is sufficient to define an elementary
  form of transient. However, the transient predicate then has to be
  extremely elaborate, typically encoding control point of the
  program, so that it is falsified by every action of the component. The other
  rules permit simpler predicates to be proven transient.

\item Basis rule is the only rule that needs program code for its
  application, others are derived from properties of the components,
  and hence, permit specification composition.


\item It is possible that $p$ is eventually falsified in every
  execution of a component though there is no proof for $\transient
  p$. To see this consider the program $f\ \fbb\ g$ in which every
  action of $f$ falsifies $p$ only if for some predicate $q$, $p\andl
  q$ holds as a precondition, and every action of $g$ falsifies $p$
  only if $p\andl \notl q$ holds as a precondition, and neither
  component modifies $q$. Clearly, $p$ will be falsified eventually,
  but this fact can not be proved as a transient property; only
  $p\andl q$ and $p\andl \notl q$ can be shown transient. As we show
  later, $p\ \lt\ \notl p$.

\end{enumerate}

\subsection{Progress Property: ensures}\label{ensures}
Property ensures for component $f$, written as $p\en q$ with
predicates $p$ and $q$, says that if $p$ holds at any moment in an
execution of $f$ then it continues to hold until $q$ holds, and
$q$ holds eventually. This claim applies even if $p$ holds after the
termination of $f$. For initial state predicate $r$, it is written
formally as $\{r\}\ f\ \{p\en q\ \bnfbar\ \}$ and defined as follows:

\infrule
 {\{r\}\ f\ \{p\andl \notl q\ \co\ p\orl q,\ \transient p\andl
     \notl q\ \bnfbar\ \}
 }
 {
 \{r\}\ f\ \{p\en q\ \bnfbar\ \}
 }

We see from the safety property in the hypothesis that once $p$ holds
it continues until $q$ holds, and from the transient property
that eventually $q$ holds.

Corresponding to each proof rule for transient, there is a similar
rule for ensures. These rules and additional derived rules for $\lren$
are given in Appendix~\ref{derived-en-rules}
(page~\pageref{derived-en-rules}).

\paragraph{Example: Distributed counter, contd.}
We prove a progress property of the annotated program from
Section~\ref{dc-contd}, reproduced below. 

\begin{gtab5}
$f_j::$\\
\tb$\initially\ old_j,\ new_j = 0,\ 0$ \\ 
\tb$\{\true\}$\\
\tb\lkwd{loop}\\
\ttb $\{\true\}$\\
\tttb $\alpha_j::\ new_j := old_j+1;$\\
\ttb $\{new_j = old_j+1\}$\\
\tttb $\lkwd{if} [\ \ \beta_j::\ \{new_j = old_j+1\}\ ctr =
  old_j\ \raa\  ctr := new_j\ \ \{\true\}$\\
\ttttb$\ \vert\ \gamma_j::\ \{new_j = old_j+1\}\ ctr \ne
old_j\ \raa\ old_j := ctr\ \ \ \{\true\}$\\
\ttttb$]$\\ 
\ttb $\{\true\}$\\
\tb\lkwd{forever}
\end{gtab5}

Our ultimate goal is to prove that for any integer $m$ if $ctr = m$ at
some point during an execution of $f$, eventually $ctr > m$. To this
end let auxiliary variable $nb$ be the number of threads
$f_j$ for which $ctr \ne old_j$. We prove the following ensures
property, (E), that says that every step of $f$ either increases $ctr$
or decreases $nb$ while preserving $ctr$'s value. Proof uses the
inheritance rule from Appendix~\ref{derived-en-rules}
(page~\pageref{derived-en-rules}).  For every $f_j$ and any $m$ and
$N$:

\begin{gtab}
$ctr = m \andl nb = N \en\ ctr = m \andl nb < N \orl ctr > m$ in $f_j$\` (E)
\end{gtab}

We use the rules for $\lren$ given in Appendix~\ref{derived-en-rules}
(page~\pageref{derived-en-rules}). First, to prove (E) in $g;\ h$, for
any $g$ and $h$, it is sufficient to show that $g$ terminates and (E)
in $h$. Hence, it is sufficient to show that (E) holds only for the
loop in $f_j$, because initialization always terminates. Next, using
the inheritance rule, it is sufficient to show that (E) holds only for
the body of the loop in $f_j$. Further, since $\alpha_j$ always
terminates, (E) needs to be shown only for the \lrkwd{if}
statement. Using inheritance, prove (E) for $\beta_j$ and
$\gamma_j$. In each case, assume the precondition $ctr = m \andl nb =
N$ of \lrkwd{if} and the preconditions of $\beta_j$ and $\gamma_j$.
The postcondition $ctr = m \andl nb < N \orl ctr > m$ is easy to see
in each of the following cases:

\begin{gtab}
$\beta_j::\ \{ctr = m \andl nb = N\andl new_j =
  old_j+1\andl ctr = old_j\}$\\
\ttttb $ \ ctr:= new_j$\\
\ttb$ \ \{ctr = m \andl nb < N \orl ctr > m\}$ \\[0.1in]
$\gamma_j::\ \{ctr = m \andl nb = N\andl new_j =
  old_j+1\andl ctr \ne old_j\} $\\
\ttttb $old_j:= ctr$\\
\ttb$ \ \{ctr = m \andl nb < N \orl ctr > m\}$
\end{gtab}

\subsection{Progress Property: Leads-to}\label{Leads-to}
The informal meaning of $p \ltt q$ (read: $p$ \lt\ $q$) is
``if $p$ holds at any point during an execution, $q$ holds
eventually''.  Unlike $\lren$, $p$ is not required to hold until $q$
holds.

Leads-to is defined by the following three rules, taken from Chandy
and Misra~\cite{CM88}. The rules are easy to justify intuitively.

\begin{itemize}
\item (basis)\quad $\inference{p\ \en\ q}
        {p\ \ltt\ q}$

\item (transitivity)\quad $\inference{p\ \ltt\ q\ ,\ q\ \ltt\ r}
        {p\ \ltt\ r}$

\item (disjunction)\quad For any (finite or infinite) set of
predicates $S\ $

\qquad \qquad \qquad $\inference{( \forall p:\ p\in S:\ p\
\ltt\ q)} {( \orl p:\ p\in S:\ p)\ \ltt\ q}$

\end{itemize}

Derived rules for $\ltt$ are given in Appendix~\ref{derived-ltt-rules}
(page~\pageref{derived-ltt-rules}). $\lt$ is not conjunctive, nor does
it obey the inheritance rule, so even if $p\ltt q$ holds in both
$f$ and $g$ it may not hold in $f\ \fbb\ g$.

\paragraph{Example: Distributed counter, contd.}
We show that for the example of Section~\ref{counter} the counter
$ctr$ increases without bound. The proof is actually quite simple. We
use the induction rule for leads-to given in Appendix
Section~\ref{Heavyweight}.

The goal is to show that for any integer $C$, $\true\ltt ctr > C$.
Below, all properties are in $f$.

\begin{gtab5}
$ctr = m \andl nb = N\ \en\ ctr = m \andl nb < N \orl ctr > m$\\
  \ttb proven in Section~\ref{ensures}\\
$ctr = m \andl nb = N\ \ltt\ ctr = m \andl nb < N \orl ctr > m$\\
  \ttb Applying the basis rule of $\lt$\\
$ctr = m \ \ltt\ ctr > m$\\
  \ttb Induction rule, use the well-founded order $<$ over
  natural numbers\\
$\true \ \ltt\ ctr > C$, for any integer $C$\\
  \ttb Induction rule, use the well-founded order $<$ over
  natural numbers.
    
\end{gtab5}

\section{Related Work } 
\HIDE{I have the following items in related work in the given order:
 UNITY, separation
logic and VCC. Would you suggest adding anything else?}

The earliest proof method for concurrent programs appears in Owicki
and Gries~\cite{OrGr76}. The method works well for small programs, but
does not scale up for large ones. Further it is limited to proving
safety properties only. There is no notion of component specification
and their composition. Lamport~\cite{Lamport} first identified $\lt$
for concurrent programs as the appropriate generalization of
termination for sequential programs (``progress'' is called
\emph{liveness} in that paper).  Owicki and Lamport~\cite{O+L} is a
pioneering paper.

The first method to suggest proof rules in the style of
Hoare~\cite{Hoare69}, and thus a specification technique, is due to
Jones~\cite{Jon83a,Jon83b}. Each component is annotated assuming that
its environment preserves certain predicates. Then the assumptions are
discharged using the annotations of the various components. The method
though is restricted to safety properties only. A similar technique
for message communicating programs was proposed in Misra and
Chandy~\cite{CM81}.

The approaches above are all based on specifying allowed interface
behaviors using compositional temporal logics. Since 2000, a number of
proposals\cite{Dinsdale-Young:2010:CAP:1883978.1884012,DBLP:conf/cav/CohenMST10}
have instead used traditional Hoare triples and resource/object
invariants, but extending state predicates to include permissions
describing how locations can be used, and suitably generalizing the
Hoare rule for disjoint parallel composition. However, these proposals
typically address only global safety and local termination properties,
not global progress properties (as this paper does).

A completely different approach is suggested in the UNITY theory of
Chandy and Misra~\cite{CM88}, and extended in Misra~\cite{seuss}. A
restricted language for describing programs is prescribed. There is no
notion of associating assertions with program points. Instead, the
safety and progress specification of each component is given by a set
of properties that are proved directly. The specifications of
components of a program can be composed to derive program
properties. The current paper extends this approach by removing the
syntactic constraints on programs, though the safety and progress
properties of UNITY are the ones used in this paper.

One of the essential questions in these proof methods is to propose
the appropriate preconditions for actions. In Owicki and
Gries~\cite{OrGr76} theory it is postulated and proved. In UNITY the
programmer supplies the preconditions, which are often easily
available for event-driven systems. Here, we derive preconditions that
remain valid in any environment; so there can be no assertion about
shared variables. The theory separates local precondition (obtained
through annotation) from global properties that may mention shared
variables.

The proof strategy described in the paper is bilateral in the following
sense. An invariant, a perpetual property, may be used to strengthen a
postcondition, a terminal property, using the invariance
rule. Conversely, a terminal property, postcondition $post_f$ of $f$,
may be employed to deduce a transient predicate, a perpetual property.

Separation logic~\cite{seplogic} has been effective in
reasonong about concurrently accessed data structures. We are studying
its relationship to the work described here.

\paragraph{Acknowledgment}
I am truly grateful to Ernie Cohen of Amazon and Jos\'{e} Meseguer of
the Univ. of Illinois at Urbana, for reading the manuscript thoroughly
and several times, endless discussions and substantive comments that
have shaped the paper. Presenting this work at the IFIPS Working Group
2.3 and the Programming Languages lunch at the University of Texas at
Austin has sharpened my understanding; I am grateful to the students
and the other attendees, in particular to Dhananjay Raju.

\newpage
\appendix
\section{Appendix: Derived Rules}

\subsection{Derived Rules for $\lrco$ and its special
  cases}\label{co-derived} 
The derived rules for $\lrco$ are given in Figure~\ref{co-rules} and for the
special cases in
Figure~\ref{derived-cox}.
The rules are easy to derive; see Chapter 5 of
Misra~\cite{seuss}.

\begin{figure*}[h]\small
\begin{tabular}{lcl}
  \hspace{-7em}

\begin{minipage}{.56\textwidth}
\infax
  {\false\ \co\  q}
  \vspace{.1in}
  
\infrule[conjunction]
  {p \ \co\  q\ ,\ p^\prime \ \co\  q^\prime}
  { p \andl p^\prime \ \co\  q \andl q^\prime}
\vspace{.1in}

\infrule[lhs strengthening]
  {p \ \co\  q}
  {p \andl p' \ \co\  q}
\vspace{.1in}
\end{minipage}
&  
\begin{minipage}{.56\textwidth}
\infax
  {p \ \co\  \true}
\vspace{.1in}

\infrule[disjunction]
  {p \ \co\  q\ ,\ p^\prime \ \co\  q^\prime}
  { p \orl p^\prime \ \co\  q \orl q^\prime}
\vspace{.1in}

\infrule[rhs weakening]
  {p \ \co\  q}
  {p \ \co\  q\orl q'}
\end{minipage}
\end{tabular}
\caption{Derived rules for $\lco$}
\label{co-rules}
\end{figure*}

The top two rules in Figure~\ref{co-rules} are simple properties of
Hoare triples.  The conjunction and disjunction rules follow from the
conjunctivity, disjunctivity and monotonicity properties of the
weakest precondition, see Dijkstra~\cite{bib:DijkstraGuarded} and of
logical implication.  These rules generalize in the obvious manner to
any set ---finite or infinite--- of \lrco-properties, because weakest
precondition and logical implication are universally conjunctive and
disjunctive.

The following rules for the special cases are easy to derive from the
definition of $\stable$, $\invariant$ and $\constant$. 
Special Cases of co

\begin{figure*}[ht]\small 

  \begin{center}
\begin{itemize}
\item (stable conjunction, stable disjunction)

\quad $\inference{p \ \co\  q\ ,\ \stable r\ \ }
        {\begin{array}{ccc}
        p \andl r \ \co\  q \andl r\\
        p \orl r \ \co\  q \orl r
        \end{array}} $

\item (Special case of the above) 
\quad $\inference{\stable p\ ,\ \stable q}
        {\begin{array}{ccc}
        \stable p \andl q\\
        \stable p \orl q
        \end{array}} $

\item \quad $\inference{\invariant p\ ,\ \invariant q\ }
        {\begin{array}{ccc}
        \invariant p \andl q\\
        \invariant p \orl q
        \end{array}} $

\item
\begin{tabular}{lcl}
\begin{minipage}{.56\textwidth}
\quad $\inference{\{r\}\ f\ \{\stable p\ \bnfbar\} }
        {\{r\andl p\}\ f\ \{p\}} $
\end{minipage}

&
\begin{minipage}{.56\textwidth}
\quad $\inference{\{r\}\ f\ \{\constant e\ \bnfbar\} }
        {\{r \andl e = c\}\ f\ \{e = c\}} $
\end{minipage}
\end{tabular}

\item (constant formation) \quad Any expression built out of constants
  is constant.

\end{itemize}
\noindent
\caption{Derived rules for the special cases of \lco}
\label{derived-cox}
\end{center}

\end{figure*}

\newpage
\subsection{Derived Rules for $\rtransient$}\label{derived-tr-rules}
\begin{itemize}
\item $\transient \false$. 

\item (Strengthening) Given $\transient p$, $\transient p\andl q$ for any $q$.

\end{itemize}

To prove $\transient \false$ use the basis rule. The proof of the
strengthening rule uses induction on the number of applications of the
proof rules in deriving $\transient p$. The proof is a template for
proofs of many derived rules for ensures and leads-to.  Consider the
different ways by which $\transient p$ can be proved in a
component. Basis rule gives the base case of induction.

\begin{enumerate}
\item (Basis) In component $f$, $p$ is of the form $p'\andl \notl
  \postf$ for some $p'$. Then in some annotation of $f$ where action
  $b\raa \alpha$ has the precondition $pre$:\\ (1) $pre\andl
  p'\ \impl\ b$, and (2) $\{pre\andl p'\}\ \alpha\ \{\notl p'\}$.\\ (1')
  From predicate calculus $pre\andl p'\andl q\ \impl\ b$, and\\ (2')
  from Hoare logic $\{pre\andl p'\andl q\}\ \alpha\ \{\notl
  p'\}$. Applying the basis rule, $\transient p'\andl q\andl
  \notl \postf$, i.e., $\transient p\andl q$.

\item (Sequencing) In $f\otw g$, $\transient p\andl \notl
  \postf$ in $f$ and $\transient p$ in $g$. Inductively, $\transient p\andl q\andl \notl
  \postf$ in $f$ and $\transient p\andl q$ in $g$. Applying the sequencing rule, $\transient p\andl q$.

\item (Concurrency, Inheritance) Similar proofs.

\end{enumerate}

\subsection{Derived Rules for $\len$}\label{derived-en-rules} 

\subsubsection{Counterparts of rules for transient}\label{transient-rules}
This set of derived rules correspond to the similar rules for
$\rtransient$.  Their proofs are straight-forward using the definition
of $\lren$.

\begin{itemize}
\item (Basis)

$\inference
 {
   \begin{array}{ccc}
     \{r\}\ f\ \{s\} \\
       \ \mbox{For every action}\  b\raa \alpha \ \mbox{with
         precondition}\  pre \ \mbox{in the annotation}:\\
       pre\andl p\andl \notl q\ \impl\ b\\
       \{pre\andl p\andl \notl q\}\ \alpha\ \{q\}
  \end{array}
 }
 {\{r\}\ f\ \{p \en p\andl s \orl q\ \bnfbar\ s\}
 }
$\\

\item (Sequencing)
\begin{gtab}  
  $\inference
 {
  \begin{array}{ccc}
    \{r\}\ f\ \{p \en p\andl \postf\orl q\ \bnfbar\ \postf\}\\
    \{\postf\}\ g\ \{p \en q\ \bnfbar\ \}
  \end{array}
 }
 {\{r\}\ f\otw g\ \{p \en q\ \bnfbar\ \}
 }
$
\end{gtab}

\item (Concurrency)
\begin{gtab}  
  $\inference
 {
  \begin{array}{ccc}
    p \en q \ \mbox{in}\  f \\
    p\andl \notl q\ \co\ p\orl q \ \mbox{in}\ g
  \end{array}
 }
 {p \en q  \ \mbox{in}\ f\ \fbb\ g
 }
$

\end{gtab}

\item (Inheritance)
Assuming the proof rule at left the inheritance proof
rule at right can be asserted. 

\begin{tabular}{lr}
\begin{minipage}{.45\textwidth}
Given:\ $\inference
 {(\forall i::\ \{r_i\}\ f_i\  \{s_i\})
 }
 {\{r\}\ f\ \{s\}
 }
$
\end{minipage}

&
\begin{minipage}{.45\textwidth}
Assert:\ $\inference
 {(\forall i::\ \{r_i\}\ f_i\ \{p\en q\ \bnfbar\ s_i\})
 }
 {\{r\}\ f\ \{p\en q\ \bnfbar\ s\}
 }
$
\end{minipage}
\end{tabular}

\end{itemize}

\subsubsection{Additional derived rules} 
The following rules are easy to verify by expanding each ensures
property by its definition, and using the derived rules for
$\rtransient$ and $\lrco$. We show one such proof, for the PSP
rule. Observe that ensures is only partially conjunctive and not
disjunctive, unlike $\lrco$.

\begin{enumerate}
\item (implication) \quad 
$\inference
 {p \impl q
 }
 {p \en q
 }
 $

Consequently, $\false \en q$ and $p\en \true$ for any $p$ and $q$.

\item (rhs weakening)\qquad \ 
$\inference
   {p \ \en\  q}
   {p \ \en\  q \orl q'} $

\item{(partial conjunction)}\qquad
  $\inference{\begin{array}{lll}
        p\ \en\ q \\
        p'\ \en\ q
        \end{array}}
  {p \andl p'\ \en\ q}$

\item (lhs manipulation)\qquad \
$\inference
   {p\andl \notl q\ \impl\ p'\ \impl\ p \orl q}
   {p \ \en\  q\ \eql\ p' \ \en\  q} $

Observe that $p \andl \notl q\ \eql\ p' \andl \notl q$ and $p \orl
q\ \eql\ p' \orl q$. So, $p$ and $q$ are interchangeable in all the
proof rules. As special cases, $p\andl \notl q \ \en\ q\ \eql\ p
\ \en\ q \eql\ p\orl q \ \en\ q$.

\item{(PSP)}
The general rule is at left, and a special case at right using $\stable r$ as $r\co r$.\\
  
\begin{tabular}{ll}
\begin{minipage}{.4\textwidth}
  \quad (PSP)\\
    $\inference{\begin{array}{lll}
        p\ \en\ q\ \\
        r \ \co\ s
        \end{array}}
  {p \andl r\ \en\ (q \andl r)\ \orl\ (\notl r \andl s)}$
\end{minipage}

&
\begin{minipage}{.6\textwidth}
  (Special case)\\
  $\inference{\begin{array}{lll}
        p\ \en\ q\ \\
        \stable r
        \end{array}}
        {p \andl r\ \en\ q \andl r}$
\end{minipage}
\end{tabular}

\item (Special case of Concurrency)
\quad

  $\inference
 {
  \begin{array}{ccc}
    p \en q \ \mbox{in}\  f \\
    \stable p\ \mbox{in}\ g
  \end{array}
 }
 {p \en q  \ \mbox{in}\ f\ \fbb\ g
 }
$ \hfill $\Box$

\end{enumerate}

\noindent
Proof of (PSP): From the hypotheses:
\begin{gtab}
  $\transient p\andl \notl q$ \` (1)\\
  $p\andl \notl q\co p\orl q$ \` (2)\\
  $r \ \co\ s$                \` (3)
\end{gtab}

We have to show:
\begin{gtab}
  $\transient p\andl r \andl \notl(q\andl r)\andl \notl(\notl r\andl s)$\` (4)\\
  $p\andl r \andl \notl(q\andl r)\andl \notl(\notl r\andl s)\co p\andl
  r\orl q\andl r\orl \notl r \andl s$ \` (5)
\end{gtab}

First, simplify the term on the rhs of (4) and lhs of (5) to $p\andl r
\andl \notl q\andl s$. Proof of (4) is then immediate, as a
strengthening of (1). For the proof of (5), apply conjunction to (2)
and (3) to get:

\Calc{
 $p\andl r \andl \notl q\ \co\ p\andl s\orl q\andl s$
\conn{\eql}{expand both terms in rhs}
 $p\andl r \andl \notl q\ \co\ p\andl r \andl s\orl p\andl \notl r
\andl s\orl q\andl r \andl s \orl q\andl \notl r \andl s$
\conn{\impl}{lhs strengthening and rhs weakening}
 $p\andl r \andl \notl q\andl s\ \co\ p\andl
  r\orl q\andl r\orl \notl r \andl s$
}

\subsection{Derived Rules for leads-to}\label{derived-ltt-rules}
The rules are taken from Misra~\cite{seuss} where the proofs are
given.  The rules are divided into two classes, {\em lightweight} and
{\em heavyweight}.  The former includes rules whose validity are
easily established; the latter rules are not entirely obvious.  Each
application of a heavyweight rule goes a long way toward completing a
progress proof. 

\subsubsection{Lightweight rules}\label{Lightweight}
\begin{enumerate}
\item (implication)\quad $\inference{p\ \impl\ q}{p\ \ltt\
q}$

\item (lhs strengthening,\ rhs weakening)\\[0.1in]\quad 
$\inference{p\ \ltt\ q}
        {\begin{array}{lll}
        p^\prime \andl p\ \ltt\ q\\
        p\ \ltt\ q \orl  q^\prime
        \end{array}}$
\item (disjunction)\quad 
$\inference{(\forall i::\  p_i\ \ltt\ q_i)}
        {(\forall i::\  p_i)\ \ltt\ (\forall i::\  q_i)}$\\

\noindent
where $i$ is quantified over an arbitrary finite or infinite index set, and $p_i, q_i$ are
predicates. 

\item (cancellation)\quad 
$\inference{p\ \ltt\ q \orl  r\ ,\ r\ \ltt\ s}
        {p\ \ltt\ q \orl  s}$
\end{enumerate}
 
\subsubsection{Heavyweight rules}\label{Heavyweight}
\begin{enumerate}
\item{(impossibility)}\quad
$\inference{p\ \ltt\ \false}
        {\notl p}$

\item{(PSP)}
The general rule is at left, and a special case at right using $\stable r$ as $r\co r$.\\
  
\begin{tabular}{ll}
\begin{minipage}{.4\textwidth}
  \quad (PSP)\\
    $\inference{\begin{array}{lll}
        p\ \ltt\ q\ \\
        r \ \co\ s
        \end{array}}
  {p \andl r\ \ltt\ (q \andl r)\ \orl\ (\notl r \andl s)}$
\end{minipage}

&
\begin{minipage}{.6\textwidth}
  (Special case)\\
  $\inference{\begin{array}{lll}
        p\ \ltt\ q\ \\
        \stable r
        \end{array}}
        {p \andl r\ \ltt\ q \andl r}$
\end{minipage}
\end{tabular}

\item{(induction)}\quad
Let $M$ be a total function from program states to a well-founded set ($W,\prec$).  
Variable $m$ in the following premise ranges over $W$. Predicates $p$
and $q$ do not contain free occurrences of variable $m$. 

\begin{gtab}
$\inference{(\forall m::\  p\ \andl\ M=m\ \ltt\ (p\ \andl\ M\prec m)\ \orl\
q)} 
{p\ \ltt\ q}$
\end{gtab}

\item{(completion)} Let $p_i$ and $q_i$ be predicates where $i$ ranges
over a finite set.

$\inference{\begin{array}{lll}
        (\forall i::\\
        \quad p_i\ \ltt\ q_i \orl  b\\
        \quad q_i \ \co\ q_i \orl  b\\
        )
        \end{array}}
        {(\forall i::\  p_i)\ \ltt\ (\forall i::\  q_i)\ \orl\ b}$
\end{enumerate}

\subsubsection{Lifting Rule}\label{leads-to Composition}
This rule permits lifting a $\lt$ property of $f$ to $f\ \fbb\ g$ with
some modifications. Let $x$ be a tuple of some accessible
variables of $f$ that includes all variables that $f$ shares with
$g$. Below, $X$ is a free variable, therefore universally
quantified. Predicates $p$ and $q$ name accessible variables of $f$
and $g$. Clearly, any local variable of $g$ named in $p$ or $q$ is
treated as a constant in $f$. 

\paragraph{(L)}
$\inference
 {
  \begin{array}{ccc}
   p\ltt q \ \mbox{in}\  f\\
   r \andl x = X \co x = X\orl \notl r \ \mbox{in}\  g
  \end{array}
 }
 {p \ltt q\orl \notl r\ \mbox{in}\  f\ \fbb\ g
 }\\
$

An informal justification of this rule is as follows. Any $p$-state in
which $\notl r$ holds, $q\orl \notl r$ holds. We show that in any
execution of $f\ \fbb\ g$ starting in a $(p \andl r)$-state $q$ or
$\notl r$ holds eventually. If $r$ is falsified by a step of $f$ then
$\notl r$ holds. Therefore, assume that every step of $f$ preserves
$r$. Now if any step of $g$ changes the value of $x$ then it falsifies
$r$ from the antecedent, i.e., $\notl r$ holds. So, assume that no
step of $g$ modifies $x$. Then $g$ does not modify any accessible
variable of $f$; so, $f$ is oblivious to the presence of $g$, and it
establishes $q$.

As a special case, we can show

\paragraph{(L')}
$\inference
{p\ltt q \ \mbox{in}\  f}
{p\andl x = M \ltt q\orl x \ne M\ \mbox{in}\  f\ \fbb\ g}\\
$

The formal proof of (L) is by induction on the structure of the proof
of $p\ltt q \ \mbox{in}\ f$. See

\noindent
{\tt
  http://www.cs.utexas.edu/users/psp/unity/notes/UnionLiftingRule.pdf}\\
for details.

\newpage
\section{Example: Mutual exclusion}\label{Mutex}
We prove a coarse-grained version of a 2-process mutual exclusion
program due to Peterson~\cite{Peterson:mutex}. The given program has a
finite number of states, so it is amenable to model-checking. In fact,
model-checking is a simpler alternative to an axiomatic proof. We
consider this example primarily because this is an instance of a
tightly-coupled system where the codes of all the components are
typically considered together to construct a proof. In contrast, we
construct a composable specification of each component and combine the
specifications to derive a proof.

\subsection{Program}
The program has two processes $M$ and $M'$. Process $M$ has two local
boolean variables, $try$ and $cs$ where $try$ remains $\true$ as long
as $M$ is attempting to enter the critical section or in it and $cs$
is $\true$ as long as it is in the critical section; $M'$ has $try'$
and $cs'$. They both have access to a shared boolean variable
$turn$. It simplifies coding and proof to postulate an additional
boolean variable $turn'$ that is the complement of $turn$,
i.e., $turn' \eql \notl turn$. 

The global initialization and the code for $M$, along with a local
annotation, is given below. The code of $M'$ is the \emph{dual} of
$M$, obtained by replacing each variable in $M$ by its primed
counterpart. Henceforth, the primed and the unprimed versions of the
same variable re duals of each other.

The ``unrelated computation'' below refers to computation
preceding the attempt to enter the critical section that does not
access any of the relevant variables. This computation may or may not
terminate in any iteration.

\begin{gtab}
  $\initially cs,cs' = \false, \false$ --- global initialization\\\\
  $M$: $\initially try = \false$\\
$\{\notl try,\ \notl cs\}$\\
  \lkwd{loop}\\
\tb --- unrelated computation that may not terminate; \\[0.1in]

\tb $\{\notl try,\ \notl cs\}$ \tsixb $\alpha$:\quad $try,\ turn := \true,\
true$;\\[0.1in]

\tb $\{try,\ \notl cs\}$ \tsixb $\beta$:\quad $\notl try' \orl turn'
\ra cs:= \true$; --- Enter critical section\\[0.1in]

\tb $\{try,\ cs\}$ \tsixb $\gamma$:\quad $try,\ cs :=
\false,\ \false$ --- Exit critical section\\
  \lkwd{forever}
  \end{gtab} 

Given that $M'$ is the dual of $M$, from any property of $M$ obtain
its dual as a property of $M'$. And, for any property of $M\ \fbb\ M'$
its dual is a property of $M'\ \fbb\ M$, i.e., $M\ \fbb\ M'$, thus
reducing the proof length by around half.

\paragraph{Remarks on the program}
The given program is based on a simplification of an algorithm in
Peterson~\cite{Peterson:mutex}. In the original version the assignment
in $\alpha$ may be decoupled to the sequence $try:= \true;\ turn:=
\true$. The tests in $\beta$ may be made separately for each disjunct
in arbitrary order. Action $\gamma$ may be written in sequential order
$try:= \false;\ cs:= \false$. These changes can be easily accommodated
within our proof theory by introducing auxiliary variables to record
the program control.

\subsection{Safety and progress properties}
It is required to show in $M\ \fbb\ M'$ (1) the safety property: both
$M$ and $M'$ are never simultaneously within their critical sections,
i.e., $\invariant \notl (cs\andl cs')$, and (2) the progress property:
any process attempting to enter its critical section will succeed
eventually, i.e., $try\ltt cs$ and $try'\ltt cs'$; we prove just
$try\ltt cs$ since its dual also holds.

\subsubsection{Safety proof: $\invariant \notl (cs\andl cs')$}
We prove below:
\begin{gtab}
$\invariant cs\impl try$ in $M\ \fbb\ M'$\` (S1)\\
$\invariant cs'\andl try\impl turn$ in $M\ \fbb\ M'$ \` (S2)
\end{gtab}

Mutual exclusion is immediate from (S1) and (S2), as follows.
\Calc{
 $cs\andl cs'$
  \conn{\impl}{from (S1) and its dual}
  $cs\andl try\andl cs'\andl try'$
  \conn{\eql}{rewriting}
 $(cs'\andl try)\andl (cs\andl try')$
\conn{\impl}{from (S2) and its dual}
 $turn\andl turn'$
\conn{\eql}{$turn$ and $turn'$ are complements}
 $\false$
}

\paragraph{Proofs of (S1) and (S2)}
First, we show the following stable properties of $M$, which
constitute its safety specification, from which (S1) and (S2) follow.

\begin{gtab}
$\stable cs\impl try$ in $M$\` (S3)\\
$\stable try\impl turn$ in $M$\` (S4)\\
$\stable cs\andl try'\impl turn'$ in $M$\` (S5)
\end{gtab}

The proofs of (S3), (S4) and (S5) are entirely straight-forward, but
each property has to be shown stable for each action. We show the
proofs in Table~\ref{mutual-safety}, where each row refers to one of
the predicates in (S3 -- S5) and each column to an action. An entry in
the table is either: (1) ``post:\ $p$'' claiming that since $p$ is a
postcondition of this action the given predicate is stable, or (2)
``unaff.'' meaning that the variables in the given predicate are
unaffected by this action execution. The only entry that is left out
is for the case that $\beta$ preserves (S5): $cs\andl try'\impl
turn'$; this is shown as follows. The guard of $\beta$ can be written
as $try'\impl turn'$ and execution of $\beta$ affects neither $try'$
nor $turn'$, so $try'\impl turn'$ is a postcondition, and so is
$cs\andl try'\impl turn'$.

\begin{table}[h]
\begin{center}
\begin{tabular}{cc|c|c|cc}
& & $\alpha$& $\beta$& $\gamma$    \\
\hline
(S3)& $cs\impl try$& post: $\notl cs$& post: $try$ & post: $\notl cs$&    \\
(S4)& $try\impl turn$& post: $turn$& unaff.: $try,\ turn$ & post:
$\notl try$&    \\
(S5)& $cs\andl try'\impl turn'$& post: $\notl cs$& see text & post: $\notl cs$&    \\
\hline
\end{tabular}
\caption{Proofs of (S3), (S4) and (S5)}
\label{mutual-safety}
\end{center}
\end{table} 

Now we are ready to prove (S1) and (S2). The predicates in (S1) and
(S2) hold initially because $\initially cs,cs' = \false,
\false$. Next, we show these predicates to be stable. The proof is
compositional, by proving each predicate to be stable in both $M$ and
$M'$. The proof is simplified by duality of the codes.

\openproof{(S1) $\stable cs\impl try$ in $M\ \fbb\ M'$:}
\begin{mproof}
$\stable cs\impl try$ in $M$                               
  \tb , (S3)  \\
$\stable cs\impl try$ in $M'$
  \tb , $cs$ and $try$ are constant in $M'$ \\
$\stable cs\impl try$ in $M\ \fbb\ M'$
  \tb , Inheritance rule   
\end{mproof}

\openproof{(S2) $\stable cs'\andl try\impl turn$ in $M\ \fbb\ M'$:} 
\begin{mproof}
$\stable try\impl turn$ in $M$                               
  \tttb , (S4)  \\
$\stable cs'\andl try\impl turn$ in $M$
  \tttb , $cs'$ constant in $M$  \\
$\stable cs'\andl try\impl turn$ in $M'$
  \tttb , dual of (S5)  \\
$\stable cs'\andl try\impl turn$ in $M\ \fbb\ M'$
  \tttb , Inheritance rule   
\end{mproof}

\subsubsection{Progress proof: $try\ltt cs$ in $M\ \fbb\ M'$}

First, we prove a safety property:
\begin{gtab}
$try\andl \notl cs\co try\orl cs$ in $M\ \fbb\ M'$\` (S6)
\end{gtab}

To prove (S6) first prove $try\andl \notl cs\co try\orl cs$ in $M$,
which is entirely straight-forward. Now variables $try$ and $cs$ are
local to $M$, therefore $\stable try\andl \notl cs$ in $M'$, and
through rhs weakening, $try\andl \notl cs\co try\orl cs$ in
$M'$. Using inheritance (S6) follows. \hfill $\Box$

Next we prove a progress property:

\begin{gtab}
$try\andl (\notl try'\orl turn')\ \ltt\ \notl try$  in $M\ \fbb\ M'$ \` (P)
  \end{gtab}

First, prove $try\andl (\notl try'\orl turn') \en \notl try$ in $M$,
using the sequencing rule for $\en$; intuitively, in a $try\andl
(\notl try'\orl turn')$-state the program control is never at
$\alpha$, execution of $\beta$ terminates while preserving $try\andl (\notl
try'\orl turn')$, and execution of $\gamma$ establishes $\notl try$.

Using duality on (S4) get $\stable try'\impl turn'$ in $M'$, i.e.,
$\stable \notl try'\orl turn'$ in $M'$. And $try$ is local to $M$, so
$\stable try$ in $M'$. Conjoining these two properties, $\stable
try\andl (\notl try'\orl turn')$ in $M'$. Apply the concurrency rule
for $\en$ with $\stable try\andl (\notl try'\orl turn')$ in $M'$ and
$try\andl (\notl try'\orl turn') \en \notl try$ in $M$ to conclude
that $try\andl (\notl try'\orl turn')\ \en\ \notl try$ in
$M\ \fbb\ M'$.

Now apply the basis rule for $\ltt$ to conclude the proof of (P).
\hfill $\Box$

\subparagraph{Proof of $try\ltt cs$ in $M\ \fbb\ M'$:} All the
properties below are in $M\ \fbb\ M'$. 

Conclude from (P), using lhs strengthening, 
\begin{gtab}
$try\andl \notl try'\ltt \notl try$ \` (P1)\\
$try\andl turn'\ltt \notl try$ \` (P2)
\end{gtab}

The main proof:

\begin{mproof}
$try'\andl turn\ltt \notl try'$
  \tb , duality applied to (P2) \\
$try\andl try'\andl turn\ltt \notl try'$
  \tb , lhs strengthening\\
$try\andl \notl (try'\andl turn)\ltt \notl try$
  \tb , rewriting (P) using $\notl turn \eql turn'$\\
$try\ltt \notl try\orl \notl try'$
  \tb , disjunction of the above two \\
$try\ltt \notl try\orl (\notl try'\andl try)$
  \tb , rewriting the rhs\\
$try\ltt \notl try$
  \tb , cancellation using (P1)\\
$try\andl \notl cs\co try\orl cs$ 
  \tb , (S6)\\
$try\andl \notl cs\ltt cs$
  \tb , PSP of above two\\
$try\ltt cs$
  \tb , disjunction with $try\andl  cs\ltt cs$
\end{mproof}

\section{Example: Associative, Commutative fold}\label{CAfold}
We consider a recursively defined program $f$ where the code of
$f_1$ is given and $f\uk$ is defined to be $f_1\ \fbb\ f_k$. This
structure dictates that the specification $s_k$ of $f_k$ must 
permit proof of (1) $s_1$ from the code of $f_1$, and (2) $s\uk$ from
$s_1$ and $s_k$, using induction on $k$. This example illustrates the
usefulness of the various composition rules for the perpetual
properties. The program is not easy to understand intuitively; it does
need a formal proof.

\subsection{Informal Problem Description}
Given is a bag $u$ on which $\oplus$ is a commutative and associative
binary operation. Define $\Sigma u$, \emph{fold} of $u$, to be the
result of applying $\oplus$ repeatedly to all pairs of elements of $u$
until there is a single element.  It is required to replace all the
elements of $u$ by $\Sigma u$. Program $f_k$, for $k \ge 1$, decreases
the size of $u$ by $k$ while preserving its fold. That is, $f_k$
transforms the original bag $u'$ to $u$ such that: (1) $\Sigma u=
\Sigma u'$, and (2) $|u| +k = |u'|$, provided $|u'| > k$, where $|u|$
is the size of $u$. Therefore, execution of $f_{n-1}$, where $n$ is
the size of $u'$ and $n > 1$, computes a single value in $u$ that is
the fold of the initial bag.

Below, $get(x)$ removes an item from $u$ and assigns its value to
variable $x$. This operation can be completed only if $u$ has an
item. And $put(x\oplus y)$, a non-blocking action, stores $x\oplus y$
in $u$. The formal semantics of $get$ and $put$ are given by the
following assertions where $u'$ is constant:
\begin{gtab}
  $\{u = u'\} \ get(z)\ \{u' = u \cup \{z\}\} $\\
  $\{u = u'\} \ put(z)\ \{u = u' \cup \{z\}\} $  
\end{gtab}

The fold program $f_k$ for all $k$, $k \ge 1$, is  given by:
\begin{gtab}
$f_1$\ttb $=\ \ |u|  > 0 \ra get(x);\ |u|  > 0 \ra get(y);\ put(x\oplus y)$\\
$f\uk$\ttb $=\ \ f_1\ \fbb\ f_k$,\quad  $k\ge 1$
\end{gtab}

Specification and proof of safety properties appear in
Section~\ref{CAfold-Safety}, next, and progress properties in
Section~\ref{CAfold-Progress}. Observe that 

\subsection{Terminal property}\label{CAfold-Safety}
The relevant safety property of $f_k$, for all $k$, $k \ge 1$, is a terminal property:
\begin{gtab}
  $\{u= u'\}\ f_k\ \{\Sigma
  u = \Sigma u',\ |u| +k = |u'|\}$
\end{gtab}

This property can not be proved from a local annotation alone because
it names the shared variable $u$ in its pre- and postconditions.  We
suggest an enhanced safety property using certain auxiliary variables.

\paragraph{Auxiliary variables}
The following auxiliary variables of $f_k$ are local to it: 

\begin{enumerate}
\item $w_k$: the bag of items removed from $u$ that are, as yet,
  unfolded. That is, every $get$ from $u$ puts a copy of the item in
  $w_k$, and $put(x\oplus y)$ removes $x$ and $y$ from
  $w_k$. Initially $w_k = \{ \}$.

\item $nh_k$: the number of halted threads where a thread halts after
  completing a $put$. A $get$ does not affect $nh_k$ and a $put$
  increments it. Initially $nh_k = 0$.

\end{enumerate}

Introduction of any auxiliary variable $aux_k$ for $f_k$ follows a
pattern: (1) specify the initial value of $aux_k$, (2) define
$aux_1$ by modifying the code of $f_1$, corresponding to
the basis of a recursive definition, and (3) define $aux_{k+1}$
in terms of $aux_1$ and $aux_k$. 

We adopt this pattern for defining $w_k$ and $nh_k$ for all $k$, $k
\ge 1$. First, the initial values of $w_k$ and $nh_k$ are $\{ \}$ and
$0$, respectively. Second, $w_1$ and $nh_1$ are defined below in
$f_1$. Third, $w_{k+1} = w_1 \cup w_k$ and $nh_{k+1} = nh_1 + nh_k$,
for all $k$. The modified program for $f_1$ is as follows where
$\langle \cdots \rangle$ is an action.

\begin{gtab}
  $f_1 = |u|  > 0 \ra \langle get(x);\ w_1 := w_1 \cup \{x\}\rangle;$\\
  \ttb  $|u|  > 0 \ra \langle get(y);\ w_1 := w_1 \cup \{y\}\rangle;$\\
  \ttb  $\langle put(x\oplus y);\ w_1 := w_1 - \{x,y\};\ nh_1 := 1\rangle$
\end{gtab}

\noindent
Note: The definition of auxiliary variables is problematic for $nh_2$,
for instance; by definition, $nh_2 = nh_1 + nh_1$. However,
these are two different instances of $nh_1$ referring to the local
variables of the two instances of $f_1$ in $f_2$. This is not a
problem in the forthcoming proofs because the proofs always refer to
indices $1$, $k$ and $k+1$, and, $nh_k$ is treated as being different
from $nh_1$.

\paragraph{Specification of safety property}
The safety specification of
$f_k$, for all $k$, $k \ge 1$, is given by:

\begin{gtab}
\tttb  $\{u = u',\ w_k = \{  \},\ nh_k = 0\}$\\
  \tsevenb\qquad  $f_k$\\[0.04in]
 $\{\constant \Sigma(u\cup w_k),\ \constant |u| + |w_k| +
  nh_k$\\
   $\ \bnfbar\ w_k = \{  \},\ nh_k = k \}$ \` (S)
\end{gtab}

\openproof{Proof of (S) for $f_1$:}
Construct the following local annotation of $f_1$.

\begin{gtab}
  \tttb $\{w_1 = \{ \},\ nh_1 = 0\}$\\[0.02in]
  $|u|  > 0 \ra \langle get(x);\ w_1 := w_1 \cup \{x\}\rangle;$\\[0.02in]
  \tttb $\{w_1 = \{x\},\ nh_1 = 0\}$\\[0.02in]  
  \tb  $|u|  > 0 \ra \langle get(y);\ w_1 := w_1 \cup \{y\}\rangle;$\\[0.02in]
  \tttb $\{w_1 = \{x,y\},\ nh_1 = 0\}$\\  [0.02in]
  \ttb  $\langle put(x\oplus y);\ w_1 := w_1 - \{x,y\};\ nh_1 := 1\rangle$\\[0.02in]
  \tttb $\{w_1 = \{ \},\ nh_1 = 1\}$
\end{gtab}

The perpetual and terminal properties of $f_1$ in (S) are easily shown using
this annotation and employing the semantics of $get$ and $put$.

\openproof{Proof of (S) for $f_{k+1}$:} by induction on $k$.
Use the following abbreviations in the proof.

\begin{gtab}
 $a_k\ \eql\  w_k = \{  \}\ \andl\ nh_k = 0$\\
 $b_k\ \eql\ \constant \Sigma(u\cup w_k),\ \constant |u| + |w_k| +
  nh_k$\\
 $c_k\ \eql\ w_k = \{  \}\ \andl\ nh_k = k$
\end{gtab}

\begin{mproof}
$\{a_1\}\  \ { f_1}\ \{b_1\ \bnfbar\ c_1 \}$                               
  \ttb , from the annotation of $f_1$\\[0.04in]
$\{a_1\}\  \ { f_1}\ \{b\uk\ \bnfbar\ c_1 \}$                               
  \ttb , $w_k$, $nh_k$ constant in $f_1$ \` (1)\\[0.04in]
$\{a_k\}\  \ { f_k}\ \{b_k\ \bnfbar\ c_k \}$                               
  \ttb , inductive hypothesis \\[0.04in]
$\{a_k\}\  \ { f_k}\ \{b\uk\ \bnfbar\ c_k \}$                               
  \ttb , $w_1$, $nh_1$ constant in $f_k$ \` (2) \\[0.04in]
$\{a_1\andl a_k\}\  f_1\ \fbb\ f_k\ \{c_1\andl c_k \}$                               
  \ttb , join proof rule on (1,2)\\[0.04in]
  $\{a_1\andl a_k\}\  f_1\ \fbb\ f_k\ \{b\uk\ \bnfbar\ c_1\andl c_k \}$
  \ttb , inheritance on (1,2)\\[0.04in]
$\{a\uk\}\ f\uk\ \{b\uk\ \bnfbar\ c\uk\}$
  \ttb , $a\uk \impl a_1\andl a_k$ and $c_1\andl c_k \impl c\uk$
\end{mproof}

The terminal property $\{u= u'\}\ f_k\ \{\Sigma u = \Sigma u',\ |u| +k
= |u'|\}$ follows from (S) as follows. The initial values of
$\Sigma(u\cup w_k)$ and $|u| + |w_k| + nh_k$ are, respectively,
$\Sigma u'$ and $|u'|$, so deduce that $\invariant \Sigma(u\cup w_k) =
\Sigma u'$ and also $\invariant |u| + |w_k| + nh_k = |u'|$.  Given
that $f_k$ is a program whose terminal properties are being proved,
we may apply the invariance rule of Section~\ref{addl-Meta-rules} with
these invariants.  So, deduce $\Sigma u = \Sigma u',\ |u| +k = |u'|$
as a postcondition of $f_k$ with the given precondition.

\subsection{Specification and proof of progress property}\label{CAfold-Progress}
The relevant progress property is that if $u$ has more than $k$
elements initially, $f_k$ terminates eventually. That is, $|u'| > k
\ltt nh_k = k$ in $f_k$. Initially $|u'| = |u|+|w_k|+nh_k$, so it is
enough to prove $|u|+|w_k|+nh_k > k \ltt nh_k = k$. Abbreviate
$|u|+|w_k|+nh_k > k$ by $p_k$ and $nh_k = k$ by $q_k$ so that for all
$k$, $k \ge 1$, the required progress property is:

\begin{gtab}
  $p_k \ltt q_k$  in $f_k$ \` (P)
\end{gtab}

The proof of (P) is by induction on $k$, as shown in
Sections~\ref{CAfold-Progress_1} and ~\ref{CAfold-Progress_k}.

\subsubsection{Progress proof, $p_1\ltt q_1$ in
  $f_1$}\label{CAfold-Progress_1} We show that $p_1 \en q_1$, from
which $p_1 \ltt q_1$ follows by applying the basis rule of $\lt$. We
reproduce the annotation of $f_1$ for easy reference.

\begin{gtab}
  \tttb $\{w_1 = \{ \},\ nh_1 = 0\}$\\[0.02in]
  $|u|  > 0 \ra \langle get(x);\ w_1 := w_1 \cup \{x\}\rangle;$\\[0.02in]
  \tttb $\{w_1 = \{x\},\ nh_1 = 0\}$\\[0.02in]  
  \tb  $|u|  > 0 \ra \langle get(y);\ w_1 := w_1 \cup \{y\}\rangle;$\\[0.02in]
  \tttb $\{w_1 = \{x,y\},\ nh_1 = 0\}$\\  [0.02in]
  \ttb  $\langle put(x\oplus y);\ w_1 := w_1 - \{x,y\};\ nh_1 := 1\rangle$\\[0.02in]
  \tttb $\{w_1 = \{ \},\ nh_1 = 1\}$
\end{gtab}

Next, prove $p_1\en q_1$ using the sequencing rule of $\lren$, from
Section~\ref{transient-rules}. It amounts to showing that if $p_1$
holds before any action then the action is effectively executed and
$q_1$ holds on completion of $f_1$. As shown in
Section~\ref{CAfold-Safety} $p_1$ is stable, and from the annotation
$q_1$ holds on completion of $f_1$. Therefore, it suffices to show
that if $p_1$ holds initially then every action is effectively
executed. The $put$ action is always effectively executed. Using the
given annotation, the verification conditions for the two $get$
actions are shown below in full:

\begin{gtab}
$w_1 = \{ \}\andl nh_1 = 0\andl |u|+|w_1|+nh_1 > 1\ \impl\ |u|  > 0$,
  and\\
$w_1 = \{x \}\andl nh_1 = 0\andl |u|+|w_1|+nh_1 > 1\ \impl\ |u|  > 0$  
\end{gtab}
These are easily proved.

\subsubsection{Progress proof, $p\uk\ltt q\uk$ in $f\uk$}\label{CAfold-Progress_k}
The main part of the proof uses the observation that every action,
$put$ and $get$, reduces a well-founded \emph{metric}. The metric is
the pair $(|u|+|w_k|,|u|)$ and the order relation is
lexicographic. Clearly, the metric is bounded from below because each
set size is at least $0$. The crux of the proof is to show that if
program $f_k$ is not terminated there is some action that can be
executed, i.e., there is no deadlock. Henceforth, abbreviate
$(|u|+|w_k|,|u|)$ by $z_k$ and any pair of non-negative integers by
$n$.

First, observe that $z_k$ can only decrease or remain the same in
$f_k$, that is, $\stable z_k \preceq n$ in $f_k$, where $\preceq$ is
the lexicographic ordering.  The proof is by induction on $k$ and it
follows the same pattern as all other safety proofs. In $f_1$,
informally, every effective $get$ preserves $|u| + |w_1|$ and
decreases $|u|$, and a $put$ decreases $|u| + |w_1|$. For the proof in
$f\uk$: from above, $\stable z_1\preceq n$ in $f_1$, and inductively
$\stable z_k \preceq n$ in $f_k$. Since $\constant w_k$ in $f_1$ and
$\constant w_1$ in $f_k$, $\stable z\uk \preceq n$ in both $f_1$ and
$f_k$. Apply the inheritance rule to conclude that $\stable z\uk
\preceq n$ in $f\uk$.

The progress proof of $p\uk\ltt q\uk$ in $f\uk$ is based on two simpler
progress results, (P1) and (P2). (P1) says that any execution starting
from $p_{k+1}$ results in the termination of either $f_1$ or
$f_k$. And, (P2) says that once either $f_1$ or $f_k$ terminates the
other component also terminates. The desired result, $p_{k+1} \ltt
q_{k+1}$, follows by using transitivity on (P1) and (P2).

\begin{gtab}
$p_{k+1} \ltt p_{k+1} \andl (q_1\orl q_k)$  in $f_{k+1}$ \` (P1)\\

$p_{k+1} \andl (q_1\orl q_k) \ltt q_{k+1}$  in $f_{k+1}$\`  (P2)

\end{gtab}

The proofs mostly use the derived rules of $\lt$ from
Section~\ref{derived-ltt-rules}.  Note that $z\uk$ includes all the
shared variables between $f_1$ and $f_k$, namely $u$, so that the
lifting rule can be used with $z\uk$. Also note that\\ $p_{k+1} \impl
(p_1 \orl p_k)$,\ $p_{k+1} \andl q_1 \impl p_k$, $p_{k+1} \andl q_k
\impl p_1$ and $q_1 \andl q_k \impl q_{k+1}$. As shown in
Section~\ref{CAfold-Safety} $p_k$ is constant, hence stable, and $q_k$
is also stable in $f_k$.

\paragraph{Proof of (P1)}  
$p_{k+1} \ltt p_{k+1} \andl (q_1\orl q_k)$ in $f_{k+1}$: Below all
properties are in $f_{k+1}$. Lifting rule refers to rule (L') of
Section~\ref{leads-to Composition}. We have already shown $p_1\ltt
q_1$ and, inductively, $p_k\ltt q_k$.

\begin{mproof2}
$p_1 \andl z\uk =n \ltt q_1 \orl z\uk \ne n$
  \tnineb , Lifting rule on $p_1 \ltt q_1 $ in $f_1$\\
$p_k \andl z\uk =n \ltt q_k \orl z\uk \ne n$
  \tnineb , Lifting rule on $p_k \ltt q_k $ in $f_k$  \\
$(p_1 \orl p_k)\andl z\uk =n \ltt (q_1 \orl q_k) \orl z\uk \ne n$\\
  \tnineb , disjunction  \\
$(p_1 \orl p_k) \andl z\uk =n \ltt (q_1 \orl q_k) \orl z\uk \prec n$\\
  \tnineb , (PSP) with $\stable z\uk \preceq n$ \\
$p_{k+1}\andl (p_1 \orl p_k) \andl z\uk =n \ltt p_{k+1}\andl (q_1 \orl q_k)
  \orl p_{k+1}\andl z\uk \prec n$\\
  \tnineb , conjunction with $\stable p_{k+1}$\\
$p_{k+1}\andl z\uk =n \ltt p_{k+1}\andl z\uk \prec n \orl p_{k+1}\andl (q_1 \orl q_k)$\\
  \tnineb , $p_{k+1} \impl (p_1 \orl p_k)$\\
  $p_{k+1} \ltt p_{k+1} \andl (q_1\orl q_k)$
  \tnineb , induction rule of $\lt$
\end{mproof2}

\newpage
\paragraph{Proof of (P2)}
$p_{k+1} \andl (q_1 \orl q_k) \ltt q_{k+1}$ in $f_{k+1}$: Below all properties are in $f_{k+1}$. 

\begin{mproof2}
$p_1 \andl z\uk =n \ltt q_1\orl z\uk\ne n$                               
  \tnineb , Lifting rule on $p_1 \ltt q_1$ in $f_1$ \\

  $p_1 \andl z\uk =n \ltt q_1\orl z\uk \prec n$                               
  \tnineb , conjunction with $\stable z\uk \preceq n$\\

$q_k \andl p_1 \andl z\uk =n \ltt q_k \andl
  q_1\orl q_k \andl z\uk \prec n$ \\
  \tnineb , conjunction  with $\stable q_k$    \\

$p_{k+1} \andl q_k \andl p_1 \andl z\uk =n \ltt p_{k+1} \andl q_k \andl
  q_1\orl p_{k+1} \andl q_k \andl z\uk \prec n$ \\
  \tnineb , conjunction with $\stable p_{k+1}$    \\

$p_{k+1} \andl q_k \andl z\uk =n \ltt p_{k+1} \andl q_k \andl z\uk \prec n
  \orl p_{k+1} \andl q_k \andl q_1$ \\
  \tnineb , $p_{k+1} \andl q_k \impl p_1$\\

$p_{k+1} \andl q_k \ltt p_{k+1} \andl q_1 \andl q_k$
  \tnineb ,  induction rule of $\lt$ \\

$p_{k+1} \andl q_k \ltt q_1 \andl q_k$
  \tnineb ,  rhs weakening \\

$p_{k+1} \andl q_1 \ltt q_1 \andl q_k$
  \tnineb , similarly  \\

$p_{k+1} \andl (q_1 \orl q_k) \ltt q_{k+1}$
  \tnineb , disjunction and $q_1 \andl q_k \impl q_{k+1}$   
\end{mproof2}


\section{References}
\bibliographystyle{plain}
\bibliography{references}

\end{document}